# A "LHC Premium" for Early Career Researchers? Perceptions from within

Tiziano Camporesi[1], Gelsomina Catalano[2,3], Massimo Florio[2], Francesco Giffoni[3]

[1] CERN European Organization for Nuclear Research

[2] Department of Economics, Management and Quantitative Methods, University of Milan

[3] CSIL Centre for Industrial Studies

This version: 04.07.2016

**Abstract**

More than 36,000 students and post-docs will be involved in experiments at the Large Hadron Collider (LHC) until 2025. Do they expect that their learning experience will have an impact on their professional future? By drawing from earlier salary expectations literature, this paper proposes a framework aiming at explaining the professional expectations of early career researchers (ECR) at the LHC. The model is tested by data from a survey involving 318 current and former students (now employed in different jobs) at LHC. Results from different ordered logistic models suggest that experiential learning at LHC positively correlates with both current and former students' salary expectations. At least two not mutually exclusive explanations underlie such a relationship. First, the training at LHC gives early career researchers valuable expertise, which in turn affects salary expectations; secondly, respondents recognise that the LHC research experience *per se* may act as signal in the labour market. Respondents put a price tag on their experience at LHC, a 'salary premium' ranging from 5% to 12% in terms of their future salaries compared with what they would have expected without such experience.

**Keywords:** Research infrastructures, Cost-Benefit Analysis, Human Capital, Expectations, Salary Premium, Large Hadron Collider

**JEL Codes**: C83, D31, D61, D81, I23, O32

**Acknowledgments**: An earlier version of this paper has been produced in the frame of the research project '*Cost/Benefit Analysis in the Research, Development and Innovation Sector*' sponsored by the EIB University Research Sponsorship programme (EIBURS), whose financial support is gratefully acknowledged. Further details on this project can be found at: http://www.eiburs.unimi.it/. Additional research work has been supported by a grant to the University of Milan by the CERN, in the frame of the Future Circular Collider Study. (https://fcc.web.cern.ch/)

**Disclaimer**: This Working Paper should not be reported as representing the views of the EIB or the CERN, or any third party. Any errors remain those of the authors.

# 1 Introduction

Early career researchers (hereafter, ECR) in physics are often involved in international collaborations (Choi et al., 2011). This is an effective way for ECR to gain practical knowledge and enhance their employment opportunities (Islam et al., 2015). Employability may be further improved when such an experience takes place in a highly renowned laboratory, such as the European Organisation for Nuclear Research (CERN), (Schopper, 2009). By analysing the career of physics students involved at the Delphi experiment at CERN's LEP electron-positron collider from 1982 to 1999, Camporesi (2001), suggests that the interest of the private sector in students and researchers who spent a period at CERN "cannot be in the knowledge of fundamental law of nature, but rather on the skills that our students acquire. […] Whatever they do go on to do, their stay at CERN certainly plays a major role" (p. 146). In the same vein, OECD (2014) emphasises that "the intellectual environment at high-energy physics (HEP) laboratories is exceptional, and is probably comparable to that of the most innovative high-technology companies" (p. 18). ECR in such an environment improve their skills by working on experiments, interacting with different cultures, writing their PhD thesis, participating to meetings, conference and workshops. These competencies can be exploited in many workplaces, even outside HEP (OECD, 2014; Boisot, 2011; Camporesi, 2001). Yet, through a survey targeted to the US High-Energy Physics community, Anderson et al. (2013) confirm that many of the skills learned in the high-energy laboratories are valued much both on academic and non-academic career path (see also Danielsson, 2013; Laurila, 2013)[1].

More in general and according to earlier literature on salary expectations, graduates have own expectations about their professional lives based on their information set (Shelley, 1994). High-energy physics increases the human capital of students by providing them technical and problem-solving capacity as well as team-work capabilities, management and communications skills. The latter have been often found poor in science graduates without such advanced experimental training (Nielsen, 2014; IOP- Institute of Physics, 2012; Sharma et al., 2008; O' Byrne et al., 2008; Rodrigues at al., 2007).

This body of research suggests that ECR salary expectations are influenced by experiential learning in international collaborations, mainly based on high-value skills acquired during such involvement. However, there is no a coherent and comprehensive explanation on why and how this relationship between research experience and rewards expectations arises. Focussing on the LHC, this paper attempts to fill this gap by answering three research questions: Does the experiential learning at LHC affect salary expectations of ECR, once controlling for their personal characteristics and other potential confounding factors? If yes, which of the acquired skills mediate the relationship between this *sui generis* experience and salary expectations? And, to what extent? How much a potential "LHC premium" is worth in terms of expected salaries?

In order to answer these questions, we analyse detailed data on 318 students and former students at LHC by using ordered logistic models. Data were collected by means of a questionnaire-based survey carried out between October 2014 and March 2015 through both on-line questionnaires and face-to-face interviews at CERN.

---

[1] Danielsson H. (2013). Students as a Bridge Building Link. CERN Presentation at NORDTEC-Conference in Lund, June. https://www.lth.se/fileadmin/lth/images/Nyhetsbilder/NORDTEK_Hans_Danielsson.pdf
Laurila, S. (2013). Students as a Bridge Building Link. Student Perspective on the Cooperation of Universities and Large Scale Research Facilities. Presentation at at NORDTEC-Conference in Lund, June. https://www.lth.se/fileadmin/lth/images/Nyhetsbilder/NORDTEK_Santeri_Laurila.pdf

The remainder of this paper is structured as follows. Section 2 describes the research methodology. Specifically, this section introduces a framework to interpret the link between a period as student or post-doc at the LHC and salary expectations. The model was drawn from the literature on salary expectations and tailored to the specific needs of this study. Section 3 presents the results. Here, the framework is tested considering both starting and end-career salary expectations. Section 4 concludes.

**2 Context and research methodology**

CERN is currently the most important particle physics laboratory in the world. It offers - mainly through the Academic institutes associated to the Experiments operating on the accelerator complex - the opportunity to spend a training period at its high-energy physics experiments mostly through two programmes addressed respectively to Bachelor and Master students and PhD students in physics, engineering and computing. Typically the undergraduate programs give the possibility to spend at CERN from 4 to a maximum of 14 months; while in the doctoral programs this period ranges from 6 to 36 months. The number of incoming students at LHC experiments was about 9,000 from 2009 (the first LHC operating year) to 2014 (CERN Personnel Statistics yearly reports). Florio at al. (2016) estimate that this number increases to 36,800 (17,400 students and 19,400 post-docs) if the 1993-2025 time span is considered.[2] It is the period that goes from the first year of construction of the LHC to the end of the first phase of operation, in 2025.

Having in mind our research questions, a structured questionnaire was drawn up as a result of extensive review of earlier literature on the topic and advice from experts as well as of CERN personnel. The survey was addressed to both current and former students at LHC. Current students are respondents who were involved in different experiments at the LHC at the time the survey was conducted. Hereafter, we refer to them simply as students. Students are mainly Bachelor, Master holders or on-going PhD students. In contrast, former students are those individuals who, after having been students at the LHC, at the time of the survey, either worked at CERN as Users, Fellows or Associates[3] or they had already left CERN and were employed in different sectors. Hereafter, we refer to them as employees.

The questionnaire was structured along four sections. The first two sections inquired about personal information and experience at LHC. They were targeted to both students and employees. Section three focused on students and it investigated on expectations about their professional career including gross starting and end-career salary expectations. The fourth section was directed to employees and inquired about both the current professional career and future expectations. Clearly, the starting salary of employees refers to their first or current professional experience and thus, it is an *observed* salary and not an *expectation*.

---

[2] The taxonomy adopted by CERN classifies students into the following categories: doctoral students (mostly from institutes participating to CERN based Experiments or directly supported by CERN for specific Applied Physics programs), CERN technical students, CERN fellows and Users. See CERN Personnel Statistics yearly reports for details. The figures reported here only refer to the apportionment of these personnel categories to the LHC; that is LHC students (doctoral and technical), LHC fellows and LHC users. Users and Fellows aged more than 35 as well as participants to summer schools or short courses are not included (see Florio et al., 2016 for details).

[3] Users are CERN's guest scientists, technicians and engineers sent to CERN as members of a visiting research team to contribute to the upgrade or analysis of experiments under a memorandum of understanding with their home institution. Fellows (Fb) are graduates of a higher educational establishment, typically with a maximum of ten years' relevant professional experience. They are appointed by the CERN for a limited period of time to perform functions within the CERN as part of their professional development. Cooperation Associates (COAS and MPAc) are scientists technicians and engineers admitted by CERN to contribute on behalf of their home institution to the execution of a collaboration under an agreement between the CERN and their home institutions (see CERN Personnel Statistics yearly reports for details, or visit http://www.useroffice.web.cern.ch)

Except for salary expectations, questions related to future outlooks utilise multiple-item constructs, measured with two different types of scale: ordinal and nominal. Ordinal scales employ five-point Likert scales, with anchors of 1 and 5, indicating the weighting assigned by individuals to a set of not mutually exclusive statements about their working experience at LHC (as examples see questions B.6 and B.7 of the questionnaire). Nominal-type scales differentiate between multiple items based on qualitative classifications such names or meta-categories. Nominal variables were coded as binary (1/0) variables. An example would be the question C.3 of the questionnaire.

Two techniques were used for statistical pre-treatment of data. The first one was factor analysis of principal components (hereafter, PCA) with Varimax rotation, applied to those variables measured by ordinal Likert scales. The suitability of the data for PCA was tested for each variable by using the KMO Measure of Sampling Adequacy with the threshold value of 0.5 (Cheung et al., 2000; Cheung and Yeung, 1998; Holt, 1997). The second technique was factor analysis of multiple correspondence (hereafter, MCA) used on nominal variables[4]. The two factor analyses were performed to homogenise the information, and obtain new, continuous variables (factor scores) to constitute the inputs for later multivariate analysis, without loss of relevant statistical information. Furthermore, the factors are linearly independent, so any possible presence of collinearity in the original data was eliminated, revealing their underlying structure.

The number of factors retained for canonical ordered logistic analysis was selected according to the criteria listed below. For PCA, the number of factors to include in later multivariate analyses was determined according to eigenvalues and cumulative percentage of variance criteria. Kaiser's (1961) rule of thumb suggests the retention of those factors with an eigenvalue greater than unity. In addition, Hair at al. (1998) suggest that in social science, factors may be stopped at least when 60 per cent of the cumulative variance was explained. In order to interpret the principal component solution, the principal component loadings were detected.

As for the MCA, it is well known that, because of the coding scheme used by the MCA to process data, the inertia (i.e. variance) of the solution space is severely underestimated (Benzécri, 1979). A better estimation of the inertia, based on pseudo-eigenvalues, was proposed by Greenacre (1993). He suggested evaluating the percentage of inertia relative to the average inertia of the off-diagonal blocks of the Burt's matrix. Thus, we made use of the Greenacre's (1993) formula to select the relevant factors (Abdi and Valentin, 2007). The interpretation of factors (i.e. dimensions) was based on their graphical projections (Greenacre 2000; Blasius and Greenacre , 1998).

We tested the influence of LHC experiential learning on both starting and end-career salary expectations by using ordered logistic regressions. One may argue that pre-career or early-stage career perceptions may not have important implications for real future career development. Actually, Van Maanen and Schein (1977) define careers as a sequence of experiences and transitions. As a result, expectations individuals form before entering in the labour market or in the entry level, do influence their making decision process about the next step during their whole professional life. This theory has been empirically validated by demonstrating that perceptions at pre-career level strongly affect subsequent salary increases (Fernandez-Mateo, 2009; Keaveny and Inderrieden, 2000). Starting and end-career salary expectations are, thus, used as dependent variables in our analysis (see also Schweitzer et al., 2014).

Drawing on contemporary research on salary expectations (Frick and Maihaus, 2016; Schweitzer et al., 2014; Maihaus, 2014) and on science (mainly, physics) graduates marketability (Islam at al., 2015; Nielsen, 2014; IOP- Institute of Physics, 2012; Jusoh et al. 2011; Hazari et al, 2010; Sharma et al., 2008), we identified the following four sets of independent variables.

---

[4] The statistical package used for this was Stata, version 13.0.

Set 1. The *personal characteristics* includes:
- *Male*. It is a dummy variable taking on the value 1 for males and 0 for females;
- *Age* is a continuous variable measured in years;
- *PhD* is a dummy variable taking on the value 1 if the highest education qualification is at least a PhD or the PhD is on-going; and 0 for master and bachelor degrees;
- *Nationality* is a dummy variable which takes on value 1 if the respondent comes from a CERN Member State and 0 otherwise[5];
- *Physics* is a dummy variable, which takes on value 1 if the academic background is physics and 0 otherwise (e.g. engineering or computer science);
- *Employee* is a dummy variable, which takes on value 1 if the respondent is an employee and 0 if he is a student.

Set 2. The *experience at LHC*. We proxied the experience at LHC with the variables listed below.
The main candidates are the type of skills acquired during that training period. Specifically, respondents were asked to what extent the following skills have improved thanks to the experiential learning at LHC (see for details, question B.7)
- *Technical skills*. It is a continuous variable (factor score) which is linked to skills such problem-solving capacity, scientific and technical skills, independent thinking, critical analysis and creativity;
- *Communication skills and leadership*. It a continuous variable (factor score) and it is related to skills such communication, team/project leadership, developing, maintaining and using networks of collaborations.

The length of the research period spent at LHC and the type of experiments respondents have worked on are also entitled proxies of the experiential learning at LHC. As a result, we propose:
- *Length of stay*. It indicates the length of the research period individuals have spent at LHC. It is a continuous variable measured in months (questions B.1 and B.2).
- *ALICE, ATLAS, CMS, LHCb*. They identify the four experiments of the LHC. Each of the experiments is codified as a dummy variable taking on the value of 1 if the respondent has worked on that experiments and 0 otherwise.

It can be argued that the longer is the stay at LHC, the more likely that ECR develop valuable skills, which in turn increase pay expectations. Using this logic, we introduce the following:

Set 3. The *moderators* includes:

---
[5] CERN Member States are Austria, Belgium, Bulgaria, Czech Republic, Denmark, Finland, France, Germany, Greece, Hungary, Italy, Netherlands, Norway, Poland, Portugal, Slovak Republic, Spain, Sweden, Switzerland, United Kingdom, and Israel.

- *Technical skills X Length of stay*. It is an interaction term between the length of the research period individuals have spent at LHC and technical skills. It is a continuous variable.
- *Communication skills and leadership X Length of stay*. It is an interaction term between the length of the research period individuals have spent at LHC and communication skills.

Set 4. The *Career-related information and perceptions* includes:
- *Application for networking*. It is a continuous variable (factor score) and it is related to the importance of networking in the decision of applying for a research period at LHC. The greater the value, the more important was for respondents to apply because of the possibility to work with world-class physicists and in a prestigious and international institution as CERN is;
- *Application for skills*. Unlike the previous variable, this factor score is linked to the relevance of developing personal and professional skills rather than to networking;
- *Salary for comparators*[6]. It is a categorical variable, which describes to what extent respondents are expected that their future salary will be higher than that earned by their peers (question C.2). It takes on value 1 if 0%, 2 if up to 10%, 3 if 11-30%, 4 if more than 30%;
- *Future sector* is a continuous variable (factor score) which is positively linked to sectors such industry, finance, and ICT and negatively related to research and university;

Summing up, we introduce a comprehensive model to test the relationships between the experience at LHC - proxied by the skills acquired and/or length of stay– (Florio et al., 2016; OECD, 2014; Boisot, 2011; Camporesi, 2001) and salary expectations, by controlling for personal characteristics (Set 1), career-related information and perceptions (Set 4) and the type of experiments, which individuals have worked on (i.e. ALICE, ATLAS, CMS, LHCb). Furthermore, we test the hypothesis according to which the predictive effect of the skill acquired at LHC and the length of stay may interact each other (Set 3) meaning that the longer is the stay at LHC, the more likely that ECR develop valuable skills, which in turn increase pay expectations. Our final point is to identify the value that ECR attach to such a working experience. To this end, we look at the marginal effects of the experiential learning spent at LHC on salary expectations. The model is shown in Figure A.1.

---

[6] We borrowed this terminology from Schweitzer et al. (2014).

**Figure 1** Analytical Framework

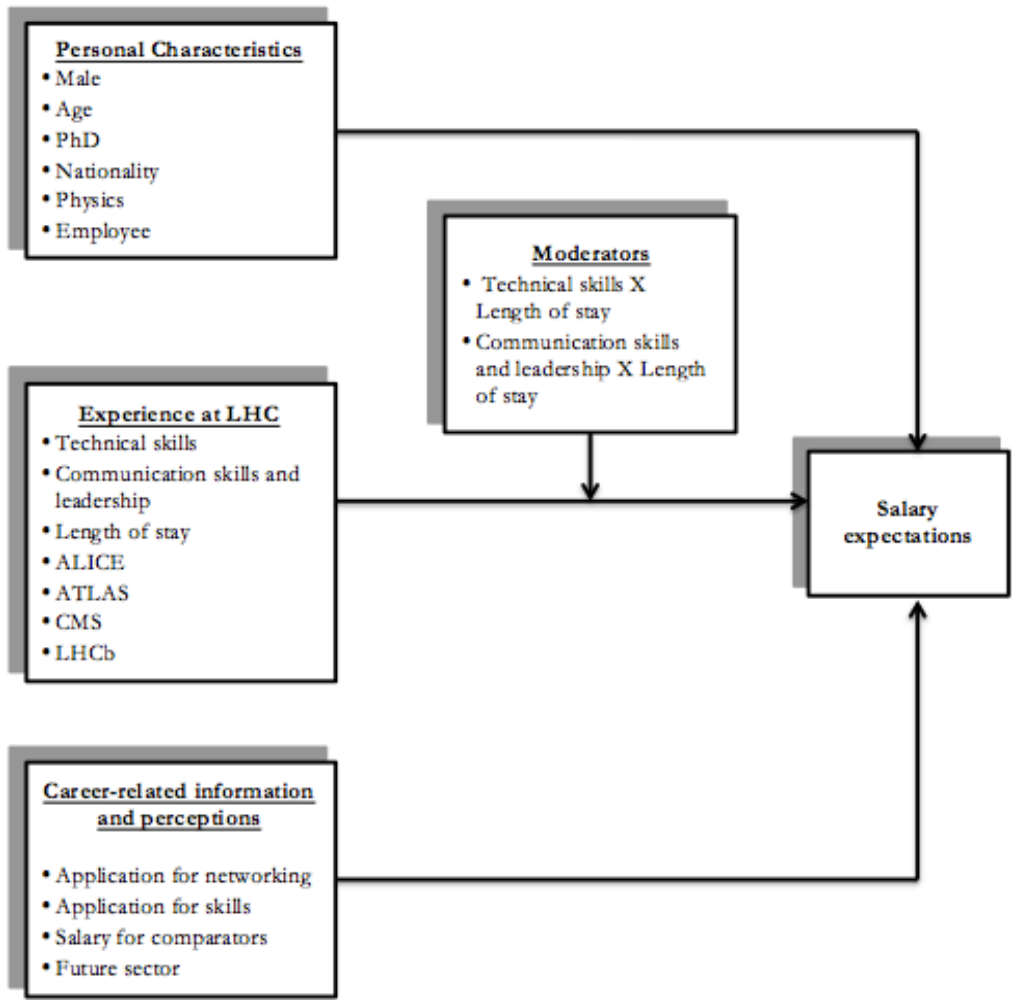

## 3 Data and Results

The study moves from data collected through a survey to current and former students working on LHC experiments coming from 52 countries (see Figure 2). The survey was carried out between October 2014 and March 2015 and resulted in 384 questionnaire collected, of which 221 through face-to-face interviews at CERN and 163 filled in online.[7] The sample used for our analysis includes 318 valid questionnaires (195 collected face-to-face and 123 online).

---

[7] The details of the survey are available in Catalano et al. (2015).

**Figure 2** Share of respondents by nationality

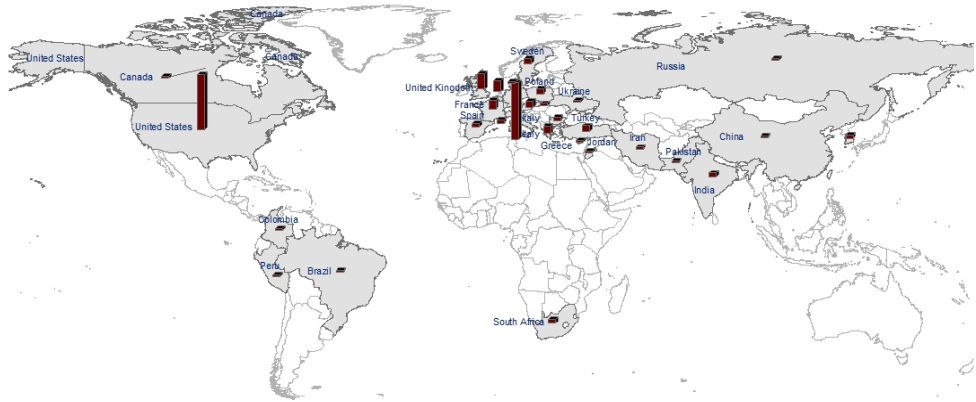

Source: *Authors processing*

Figure A.1 shows that respondents are mostly European, with an overall representation of 52 nationalities. Respondents from Italy and the USA account for the largest share (38%) followed by the UK and Germany (8%, each).

Table 1 reports the descriptive statistics of personal characteristics of respondents and the length of period they have spent at LHC. Males represent 73% of the sample (71% amongst students and 75% amongst employees). 71% of respondents have at least a PhD as their highest education level; the remainder are bachelor or master degree holders. Amongst the employees, the percentage of those with a PhD is 90%. Actually, they are mostly post-docs.

**Table 1** Personal characteristics and length of period at LHC.

| Variable | Total (*n*=318) | Students (*n*=141) | Employees (*n*=177) |
|---|---|---|---|
| **Discrete Variables** | | | |
| Gender (%) | | | |
|   Male | 73.3 | 70.9 | 75.1 |
|   Female | 26.7 | 29.1 | 24.9 |
| | | | |
| Education (%) | | | |
|   At least PhD | 71.4 | 48.2 | 89.9 |
|   Less than PhD | 28.6 | 51.8 | 10.1 |
| | | | |
| Nationality (%) | | | |
|   Member State | 62.3 | 61.7 | 63.3 |
|   Non-Member State | 37.4 | 38.3 | 36.7 |
| | | | |
| Academic background (%) | | | |
|   Physics | 85.5 | 80.1 | 89.3 |
|   Other | 14.5 | 19.1 | 10.7 |
| **Continuous Variables** | | | |
| Age (years) | | | |
|   Mean | 31.1 | 28.2 | 33.4 |
|   Std. Dev. | 4.7 | 3.5 | 4.0 |
|   Min | 21 | 21 | 25 |
|   Max | 44 | 38 | 44 |
| | | | |
| Length of stay at LHC (months) | | | |
|   Mean | 44.7 | 24.4 | 59.3 |
|   Std. Dev. | 34.7 | 17.4 | 36.6 |
|   Min | 1 | 1 | 1 |
|   Max | 181 | 72 | 181 |

With regard to the academic background, 85% are physicists, while the remaining 15% have a degree in engineering or computer sciences (a more detailed breakdown is shown in Figure 3).

**Figure 3** Share of respondents by educational degree and academic background

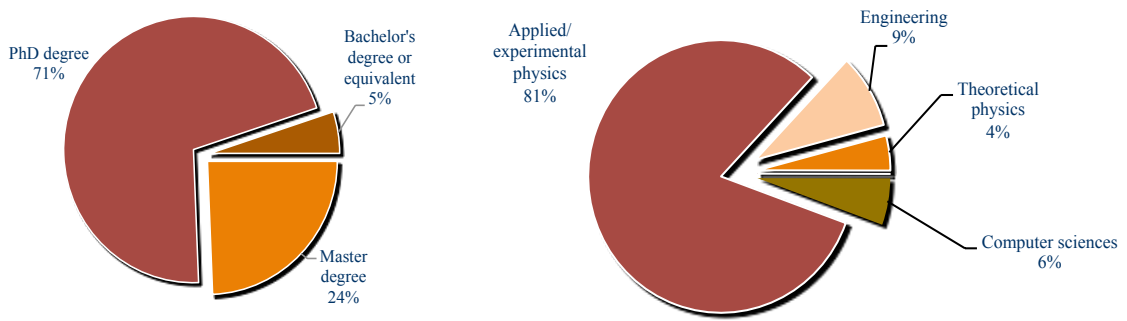

About 63% of the sample is from a CERN Member State. The average age of respondents is 31 years with an average of 45 months working experience at LHC. Among students (average age equal to 28 years) the average training period is 24 months, while among employees (average age equal to 33 years), the average length of stay at LHC is about 60 months. The excessive length of stay recorded among employees with respect to students, indicates that in our sample there is a fairly large share of employees currently working at CERN as Users, Fellows or Associates (Figure 4). Finally, the distribution of respondents among the different experiments is as follows: 5% at ALICE, 25% at ATLAS, 62% at CMS and 8% at LHCb (Figure 5). Note that in our sample, ECR working on CMS are over-represented because the survey was first launched to CMS current and former students in collaboration with the CMS team[8]. Afterwards, the survey was extended to other experiments as well.

**Figure 4** Employment sector. Share of employees   **Figure 5** Distribution of respondents across LHC

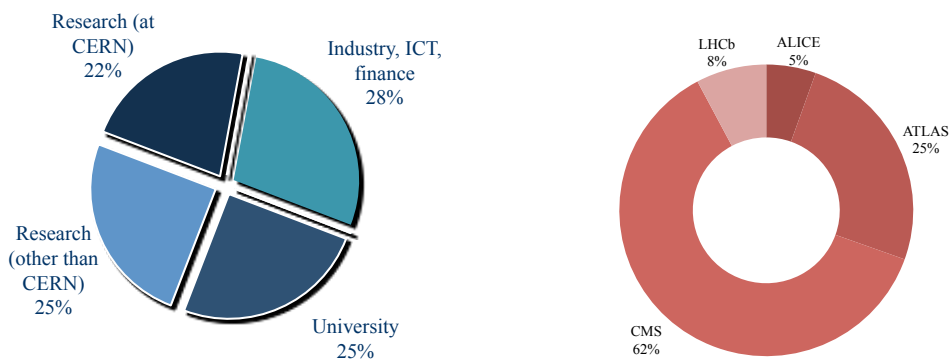

Figure 6 and Figure 7 report, respectively, descriptive statistics of the variable related to skills acquired at LHC (question B.7) and the kind of activity on which respondents have spent most of time during such an experience (question B.5).

---

[8] The support in gathering data and contacts of CMS current and former students is gratefully acknowledge to Tiziano Camporesi (CMS international coordinator) and Nicoletta Barzaghini (CMS Secretariat).

**Figure 6** Skills improved thanks to the LHC experience

**Figure 7** Distribution of time spent across activities

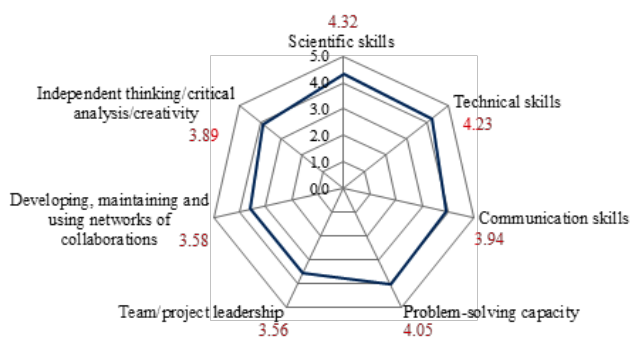
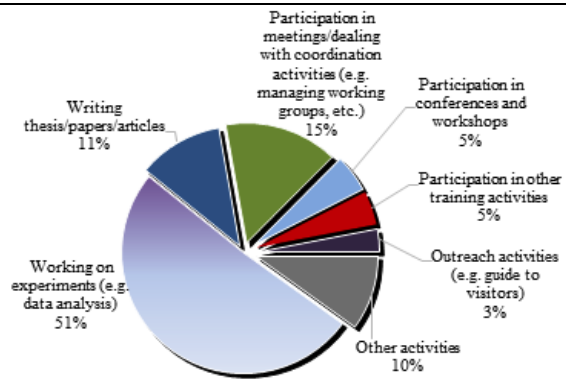

Figure 6 displays that, according to respondents, the LHC experience has improved their technical skills more than communication and leadership skills; while Figure 7 shows that most of the time respondents have spent at LHC, was dedicated to working on experiments, and specifically, data analysis (51%) and writing papers and/or thesis (11%).

The distribution of starting and end-career gross salary expectations split by employment status is reported in Table 2.

**Table 2** Gross salary expectations distributions. Percentage.

| Category | Starting-career salary | | End-career salary | |
|---|---|---|---|---|
| | Students (%) | Employees (%) | Students (n) | Employees (%) |
| < 30,000 EUR | 22.7 | 13.1 | 4.3 | 2.3 |
| 30,000-40,000 EUR | 24.1 | 23.2 | 5.1 | 1.7 |
| 40,000-50,000 EUR | 24.1 | 14.9 | 3.6 | 5.2 |
| 50,000-60,000 EUR | 10.1 | 16.7 | 19.6 | 11.6 |
| > 60,000 EUR | 19.1 | 32.1 | 67.4 | 79.2 |

**Note**: Employment status differences in the expected salaries were assessed using a Pearson's chi-square test for starting-career salary distribution and a Fisher's exact test for end-career salary distribution. Both tests reject the null of the similarity of distributions.

It shows that responses about initial career expected salaries tend to group in the lowest salaries categories (less than EUR 50,000). This particularly holds among students. For employees, the distribution of initial career salary is quite heterogeneous among categories. In contrast, the distribution of peak-career salary expectations is concentrated in the highest categories (more than EUR 50,000) for both students and employees. In order to test whether students and employees differ in their expected salaries, we carries out a Pearson's chi-square test in the initial salary case and a Fisher's exact test in the case of end career salaries expectations[9]. The chi-square test ($p <0.01$) and the Fisher's exact test ($p <0.05$) suggest that there is a statistically significant difference between students and employees in expected salaries. We control for such dissimilarity in the following multivariate analysis by including an employment status dummy variable.

---

[9] The use of two different tests is necessary because the chi-square test assumes that the value of each cell is five or higher. While this assumption is met in the distribution of starting salary expectations, it does not hold for expected salary at peak of careers. Indeed, many tapes in the lowest categories are less than five.

Table 3 analyses the overall correlation between the variables entering in the analytical framework[10]. Statistically significant correlations are observed between salary expectations (both starting and end-career expectations) and *Technical Skills* and *Length of stay* respectively; in contrast, *Communication skills/leaderships* do not correlate with salary expectations. Personal characteristics such as *Male* and *Employees* and career-related perceptions such as *Salary for comparators* and *Future sector* positively correlate with salary expectations as well.

We answer our research questions using ordered logistic models (Long and Freese, 2014). In principle, we may use both starting salary expectations and end-career salary expectations as dependent variables (Schweitzer et al., 2014). Actually, it is worth noting that, in our sample, they are strongly and positively correlated (*coef = 0.62, p <0.05;* Table 3) suggesting that using two different regression analysis wouldn't lead us to notably different conclusions. In addition, European Commission (2014, Chapter 7) suggests that the benefit of human capital development should be measured on the lifelong salary. Therefore, we only make use of end-career salary expectations as dependent variable[11]. Results are shown in Table 4. For each of the regression proposed, the proportional odds assumption, underlying ordered logistic procedure, was tested (Long and Freese, 2014, Chapter 7)[12]. The p-values are reported in the last row of the table.

We carried out the analysis in five steps. In the first step (Column 1) we only include the types of skills respondents declared having improved thanks the training at LHC, which is one of our proxy of the experiential learning. Actually, in the regressions, we only included the variable (factor score) *Technical skills*; the variable *Communication skills and leadership* was never found statistically significant. In the second step (Column 2), we test the length of the research period as proxy of training at LHC; in doing so, we exclude *Technical skills* and include *Length of stay*. The third step (Column 3) shows that *Technical skills* and *Length of stay* remain significantly associated with salary expectations also when both variables are jointly plugged into the same model. One may argue that skills acquired at high-energy physics experiments increase or improve as the length of the research period increases. We test this hypothesis in the fourth step (Column 4) by adding the so-called *moderators*. The fifth step (Column 5) presents the whole model, which controls for personal characteristics, career-related information and perception of respondents. We also tried to include in the analysis, others interaction terms to allow more hypotheses to be tested. We failed to find any statistical evidence on the contribution of such interaction terms on salary expectations. Thus, they are not reported in Table 4[13]. Regardless the step, we always control for four types of specific-effects: first, the employment status (employee versus student). It enables us to capture unobserved heterogeneity that may shape salary expectations of such individuals beyond the experience at LHC.

---

[10] The variables associated to the type of experiment ALICE, ATLAS, CMS and LHCb do not show any significant correlation with the relevant variables we are interested in (i.e. salary expectations, technical skills, communication skills/leadership and length of stay). Thus, they are not reported in the table.

[11] Regressions, which make use of starting-career salary expectations as dependent variable are available upon request. However, results mirror the ones presented in Table 4.

[12] One of the assumptions underlying ordered logistic regression is that the relationship between each pair of outcome groups is the same. Put differently, the proportional odds assumption requires that the coefficients describing the relationship between, let's say, the lowest versus all higher categories of the response variable are the same as those that describe the relationship between the next lowest category and all higher categories, etc. Because the relationship between all pairs of groups is the same, there is only one set of coefficients (only one model); otherwise, a generalized ordered logistic model should be run. In order to test the proportional odds assumption, we run the Brant test, rather then the "omodel" command, since the latter does not recognized categorical variables. The null hypothesis is that there is no difference in the coefficients between models. In our case, the proportional odds assumption is met in all of proposed regressions, except for the first model (Column 1, Table 3). For further details see Long and Freese (2014) or visit http://www.ats.ucla.edu/stat/stata/dae/ologit.htm.

[13] In an unreported regression, several interaction terms between personal characteristics and career-related information and our proxies of experience at LHC (Technical skills and Length of stay) were tested as suggested by Hogue et al. (2010). We found no statistical significance. However, results are available upon request.

**Table 3** Correlations Matrix

| | 1 | 2 | 3 | 4 | 5 | 6 | 7 | 8 | 9 | 10 | 11 | 12 | 13 | 14 | 15 |
|---|---|---|---|---|---|---|---|---|---|---|---|---|---|---|---|
| 1. Starting career salary | 1 | | | | | | | | | | | | | | |
| 2. End career salary | 0.62** | 1 | | | | | | | | | | | | | |
| *Personal characteristics* | | | | | | | | | | | | | | | |
| 3. Male | 0.23** | 0.19** | 1 | | | | | | | | | | | | |
| 4. Age | 0.19** | 0.02 | 0.11** | 1 | | | | | | | | | | | |
| 5. PhD | 0.08 | 0.04 | 0.01 | 0.33** | 1 | | | | | | | | | | |
| 6. Nationality | -0.08 | -0.07 | -0.08 | -0.03 | -0.11** | 1 | | | | | | | | | |
| 7. Physics | -0.09 | -0.14** | -0.05 | -0.20** | 0.40** | -0.17** | 1 | | | | | | | | |
| 8. Employee | 0.18** | 0.12** | 0.04 | 0.56** | 0.46** | 0.02 | 0.12** | 1 | | | | | | | |
| *Experience at LHC* | | | | | | | | | | | | | | | |
| 9. Technical skills | 0.10** | 0.12** | -0.07 | 0.00 | 0.07 | -0.03 | 0.00 | 0.02 | 1 | | | | | | |
| 10. Communication skill/leadership | -0.01 | -0.05 | -0.01 | 0.20** | 0.10 | 0.00 | -0.04 | 0.15** | -0.01 | 1 | | | | | |
| 11. Length of stay | 0.20** | 0.14** | 0.01 | 0.40** | 0.33** | 0.14** | 0.12** | 0.50** | 0.12** | 0.05** | 1 | | | | |
| *Career-related information and perceptions* | | | | | | | | | | | | | | | |
| 12. Application for networking | -0.03 | 0.02 | -0.11** | 0.01 | -0.08 | 0.16** | -0.09 | 0.00 | 0.00 | 0.38** | -0.07 | 1 | | | |
| 13. Application for skills | 0.02 | -0.00 | -0.08 | -0.04 | 0.01 | 0.02 | 0.10 | -0.14** | 0.37** | -0.20** | -0.10** | 0.29** | 1 | | |
| 14. Salary for comparators | 0.11** | 0.15** | -0.01 | -0.07 | -0.03 | -0.04 | -0.12** | -0.03 | 0.16** | -0.03 | -0.02 | 0.18 | 0.10 | 1 | |
| 15. Future sector | 0.15** | 0.18** | 0.10 | -0.01 | -0.01 | 0.01 | -0.09 | 0.12 | -0.00 | -0.13** | 0.12 | -0.16** | -0.19** | 0.03 | 1 |

**Notes:** The variables associated to the type of experiment ALICE, ATLAS, CMS and LHCb do not show any significant correlation with the relevant variables we are interested in (i.e. salary expectations, technical skills, communication skills/leadership and length of stay). Thus, they are not reported in the table. ** Significant at 5% level.

Second, nationality-fixed effects. To the extent that individuals form their salary expectations according to some features of the country of origin, for example labour market conditions or the prevailing type of educational system (Economist, 2016; Maihaus, 2014; Jusoh at al., 2011; Wickramasinghe and Perera, 2010; Hazari et.al, 2010), this dummy should capture such an effect. Third, experiments-specific effects. These dummies identify the experiments at which respondents have spent their training period at LHC: ALICE, ATLAS, CMS and LHCb[14]. Last but not least, interview-specific effect.[15] It allows us to reduce any systematic difference between responses obtained by hand and through online questionnaire (Duffy at al., 2005).

Column 1 and Column 2 reveal that experience at LHC positively and significantly correlates with salary expectations both when it is proxied by the acquired competences and by *Length of stay*. Coefficients are respectively 0.103 and 0.009 significant at 10% level. These variables keep their statistically significance up also when they are plugged simultaneously into the same model (Column 3, *coef = 0.110* and *0.009, respectively, p <0.10*), suggesting that the time spent at LHC generates *per se* increasing salary expectations, aside the skills acquired.

Column 4, adds the interaction term between *Technical Skills* and *Length of stay*. The positive and statistically coefficient on the interaction term (*coef = 0.004, p <0.05*) indicates that the skills acquired at LHC increases as the time spent on the experiments increases, which in turn generates higher rewards expectations. This is confirmed by the fact that the variable *Technical Skills* loses its predictive power in explaining salary expectations. As before, *Length of stay* retains its own significance (*coef = 0.016, p <0.01*).

The estimated association between salary expectations and experiential learning at LHC remains robust also after adding respondents' personal characteristics as well as their career-related information and perceptions (Column, 5). In addition, the coefficient on Male is positive and statistically significant at 1 percent level reflecting a substantial gender gap in salary expectations among graduates (Schweitzer et al., 2014; Hogue et al., 2010; Ng and Wiesner, 2007), and, particularly among physicists (Hazari et.al, 2010; IOP- Institute of Physics, 2012; Lissoni et al., 2011). The variable PhD enters positively and significantly as well, confirming that salary expectations increase with educational attainment (Islam at al., 2015; Jusoh et al., 2011; Shelley, 1994). There are no significant differences on end-career salary expectations between employees and students; this result means that once controlling for personal characteristics, the employment status loses its predictive power in explaining end-career expected salaries. This is probably due to the fact that, after all, the community of HEP is relatively small and information circulates amongst ECR of different seniority, at least for not too distant cohorts.

Column 5 also shows that the variables *Salary for comparators* and *Future sector* enter into the model with a significant and positive coefficient. The former variable suggests that the higher the salary respondents are expected to earn with respect to their comparable peers thanks to their research experience at LHC, the higher their own salary expectations are (Schweitzer et. al., 2014). As regard *Future sector,* higher salaries are expected in sectors such industry and finance; in contrast, respondents are expected lower salaries in academia.

---

[14] Even though these dummies variables could be potential interesting for the purpose of our analysis, we found them never statistically significant.
[15] This dummy variable takes on the value of 1 if the interview was carried out face-to face and 0 otherwise.

**Table 4** Ordered logistic estimates. Dependent variable is End Career Salary Expectation.

| Variables | (1) Coef | se | (2) coef | Se | (3) coef | Se | (4) coef | se | (5) coef | se |
|---|---|---|---|---|---|---|---|---|---|---|
| ***Experience at LHC*** | | | | | | | | | | |
| Technical skills | 0.103* | (0.062) | | | 0.110* | (0.061) | 0.004 | (0.145) | 0.135 | (0.134) |
| Length of stay | | | 0.009* | (0.005) | 0.009* | (0.005) | 0.011** | (0.005) | 0.017** | (0.007) |
| Technical skills X Length of stay | | | | | | | 0.004** | (0.002) | 0.004** | (0.002) |
| ***Personal Characteristics*** | | | | | | | | | | |
| Employee | 0.814*** | (0.282) | 0.455 | (0.346) | 0.493 | (0.352) | 0.500 | (0.354) | 0.444 | (0.409) |
| Male | | | | | | | | | 0.946*** | (0.349) |
| Age | | | | | | | | | -0.035 | (0.043) |
| PhD | | | | | | | | | 2.653*** | (0.924) |
| Physics | | | | | | | | | -0.294 | (0.449) |
| ***Career-related information*** | | | | | | | | | | |
| Application for networking | | | | | | | | | -0.098 | (0.157) |
| Application for skills | | | | | | | | | 0.272 | (0.239) |
| Salary for comparators | | | | | | | | | 0.342*** | (0.130) |
| Future sector | | | | | | | | | 0.495*** | (0.155) |
| | | | | | | | | | | |
| Nationality-specific effects | Yes | | Yes | | Yes | | Yes | | Yes | |
| Experiments-specific effects | Yes | | Yes | | Yes | | Yes | | Yes | |
| Interview-specific effects | Yes | | Yes | | Yes | | Yes | | Yes | |
| Observations | 318 | | 318 | | 318 | | 318 | | 318 | |
| McFadden's R2 | 0.036 | | 0.035 | | 0.043 | | 0.050 | | 0.159 | |
| Log Likelihood | -254.3 | | -240.8 | | -237.4 | | -235.9 | | -172.8 | |
| Likelihood ratio test | 16.87 | | 17.99 | | 19.17 | | 22.75 | | 52.20 | |
| Proportional odds hp test (p-value) | 0.291 | | 0.276 | | 0.227 | | 0.205 | | 0.182 | |

Table shows the determinants of the probability of falling in one of the expected salary category. Robust standard errors in parentheses. ***, **, * denote significance at the 1%, 5% 1% level respectively.

Finally, the likelihood ratio tests in the models indicate that the variation in the independent variables explains a good proportion of the variability in the response variable[16]. In order to assess the "LHC premium", we look at marginal effects of the working experience at LHC (proxied by the *Length of stay*) on end-career salary expectations. If a premium is expected, then it should be measured on end/peak salary expectations (Florio et al., 2016; Schweitzer et al., 2014; European Commission, 2014, Chapter 7). Marginal effects are those stemming from the full model (Column 5, Table 4) and they are shown in Table 5, Column (A); Column (B) reports the values in percentage terms.

**Table 5** Marginal effects of *Length to stay* on End-career salary expectations

| End-career salary expectations categories | (A) marginal effects | se | (B) marginal effects (%) | se |
| --- | --- | --- | --- | --- |
| < 30,000 EUR | -0.00066* | (0.00035) | -0.066* | (0.035) |
| 30,000-40,000 EUR | -0.00038* | (0.00021) | -0.038* | (0.021) |
| 40,000-50,000 EUR | -0.00039* | (0.00021) | -0.039* | (0.021) |
| 50,000-60,000 EUR | 0.00125** | (0.00057) | 0.125** | (0.057) |
| > 60,000 EUR | 0.00268*** | (0.00108) | 0.268*** | (0.108) |

**Notes:** Ma ***,**,* denote significance at the 1%, 5% 10% level respectively.

As expected, one additional month of training spent at LHC increases the probability of declaring an expected salary falling in the two highest categories (50,000-60,000 EUR and >60,000 EUR) and reduces the probability of expecting a low salary (less than 50,000 EUR). For example, an additional month of experiential learning at LHC lowers the likelihood of declaring an expected salary less than 30,000 EUR by about 0.07 percentage points; in contrast, it increases the probability of expecting a salary greater than 60,000 EUR by 0.27 percentage points, ceteris paribus.

Let's now focus on the two highest categories (*coef = 0.00125 and 0.00268, p <0.05; p <0.01*), which contain almost 85% of responses. Note also that, in our sample, the average number of months spent at LHC is 44 for the whole sample, 24 for students and 60 for employees. Thus, for an "average" individual who declared an expected salary between EUR 50,000 and 60,000 the experiential learning at LHC is worth about 5% (3% for a student and 7% for an employee). For those respondents whose expected salary falls in the category ">60,000", the stay at LHC is worth, on average, about 12% (6% for a student and 16% for an employees).

## 4 Conclusions

This paper contributes to the literature on experiential learning and salary expectations by performing for the first time a statistical analysis of the perceived professional premium of early career researchers in a high-energy physics laboratory. Survey data have been collected from students and former students at the largest particle accelerator in the world, the Large Hadron

---
[16] For the sake of simplicity, we chose to not include the constants of the regressions in Table 4. In the ordered logistic models, the constants (here, we have four constants for each model) are cut-points used to differentiate the adjacent levels of the dependent variable. Apart from some exceptions in Columns 2 and 3, they were found all statistically significant, justifying the use of five categories of the level of salary expectations over combining some categories. Actually, some preliminary elaboration on original data leads us to reduce the salary expectation categories from ten (see questions C.7 and C.10 in the questionnaire) to five.

Collider. We were particularly interested to understanding to what extent earlier results from Camporesi (2001) on students involved in experiments at LEP, a previous major collider at CERN, are confirmed for more recent cohorts of early career researchers. Moreover, and this is the novelty of our contribution, we wanted to focus on the econometric estimation of a perceived salary premium of the learning experience at the LHC and of the drivers of such perceptions.

There are several reasons why this context is of interest for a broader research perspective on professional expectations. The LHC operates at the frontiers of science, and for this reason it attracts students from a very large number of countries (over 50 in our sample). This fact ensures that possible specific country effects play a minor role. The research community of particle physics can be considered as a relatively small but dense global social network, where information on career opportunities is widely shared within each cohort and across cohorts of early career researchers. Moreover, there is fragmentary but interesting evidence that ERC at CERN will have a professional future in a variety of jobs, beyond academic research, including in industry and finance. Thus, it seems that the LHC context, including its experiments, such as CMS and ATLAS, is an ideal testing ground for the more general question of the experiential learning effect on salary expectations.

There are three main findings of our econometric analysis of survey data. First, there is no statistical difference in end-career salary expectations between the two sub-samples of respondents: current students and former students. In fact the latter, who have acquired more direct information, are on average more optimistic in their perceptions of the salary premium, but the difference is not statistically significant after controlling for individual characteristics. This suggests that the research community actually shares the information on professional opportunities and this fact shapes homogeneous expectations. This is also indirect evidence of realism of the expectations, because for former students they are based on actual information.

A second finding is that the core drivers of the expectations are *length of stay* at the LHC and *technical skills* acquired. The perceived professional premium is not a purely reputational effect in the job market associated with the mere fact of having been selected for training at CERN, but it increases proportionally to the time spent in research in that context. Respondents were able to indicate on a five-point scale which were the most important skills acquired, and we find that the salary premium increases proportionally to the perceived importance of technical skills. This result clearly points to the perception of experiential learning as a driver of professional opportunities. Finally, the interaction between the two drivers is statistically significant.

Overall, we conclude that, according to the convergent perception of respondents, either current or former students at CERN ( the latter now employed in a variety of occupations), there is a professional premium arising from sustained experiential learning in the laboratory. The premium is estimated in the range 5-12% over the entire career, a non negligible effect given the number of students involved over the LHC life. In other words, the laboratory environment is at the same time perceived by insiders as a scientific discovery machine and an engine of human capital formation, and for the first time we have been able to measure this effect from within.

# Annex I – Questionnaire

| PART A – PERSONAL INFORMATION |||||
|---|---|---|---|---|
| A.1 | *Are you:* | | Male | Female |
| A.2 | *Year of birth:* | | | |
| A.3 | *Your nationality* | | | |
| A.4 | *Your highest educational qualification* | Bachelor's degree or equivalent <br> Master degree <br> □ PhD student <br> □ Other, please specify: |||
| A.5 | *Please, indicate the institution where you have attained your highest educational degree* | |||
| A.6 | *Please, indicate the academic background of your highest educational degree* | □ Theoretical physics <br> □ Applied/experimental physics <br> □ Engineering <br> □ Computer sciences <br> □ Mathematics <br> □ Life science <br> □ Other, please specify: |||
| PART B – YOUR EXPERIENCE AT LHC |||||
| *In answering questions from B.0 to B.4, please refer to your first experience at LHC. If you have gained further experiences (e.g. working on two different experiments or other experiments not related to LHC, please indicate this information in question B.10 and specify your activity (e.g. fellow, postdoc, etc.)* |||||
| B.0 | *Please, indicate on which of the following activities you have been working when you started your experience at LHC* | □ ATLAS experiment <br> □ CMS experiment <br> □ ALICE experiment <br> □ LHCb experiment <br> LHC machine <br> HiLumi LHC <br> LIU (Injector Upgrade) <br> □ Other activities related to LHC, please specify |||
| B.1 | *Start date of your research period at LHC* | Month: <br> Year: |||
| B.2 | *End date of your research period at LHC* | Month: <br> Year: <br> □ On-going → *Expected to be completed in:* |||

| | | | | | | | |
|---|---|---|---|---|---|---|---|
| B.3 | **Please, indicate your educational qualification when you started your research period at LHC** | ☐ Bachelor's degree or equivalent<br>☐ Master degree<br>☐ PhD student<br>☐ Other, please specify | | | | | |
| B.4 | **What is/was your university affiliation during the research period at LHC?** | | | | | | |
| B.5 | **Considering that your time spent at LHC is equal to 100%, please indicate the percentage dedicated to the following activities:**<br><br>*If the allocation of the time addressed to these activities has changed during your stay at LHC, please indicate an average percentage.* | Working on experiments (e.g. data analysis) | | | | | |
| | | Writing thesis/papers/articles | | | | | |
| | | Participation to meetings/dealing with coordination activities (e.g. managing working groups, etc.) | | | | | |
| | | Participation to conferences and workshops | | | | | |
| | | Participation to other training activities | | | | | |
| | | Outreach activities (e.g. guide to visitors) | | | | | |
| | | Other activities, please specify: | | | | | |

| | | | 1 | 2 | 3 | 4 | 5 |
|---|---|---|---|---|---|---|---|
| B.6 | **How do you rate the importance of the following considerations on your decision of applying for a research period at LHC?**<br>*Please, provide a rate from 1 (= not important) to 5 (= very important) to each of the following items by keeping in mind that they are not mutually exclusive:* | Deepening the knowledge and competences in the scientific domain of interest | | | | | |
| | | Develop new professional skills | | | | | |
| | | World undisputed prestige of CERN | | | | | |
| | | Possibility to work with world class physicists | | | | | |
| | | Working in an international environment | | | | | |
| | | Other, please specify: | | | | | |

| | | | 1 | 2 | 3 | 4 | 5 |
|---|---|---|---|---|---|---|---|
| B.7 | **To what extent the following skills have been improved thanks to the experience at LHC?**<br>*Please, provide a rate from 1 (= not decisive) to 5 (= very decisive) to each of the following items by keeping in mind that they are not mutually exclusive:* | Scientific skills | | | | | |
| | | Technical skills | | | | | |
| | | Communication skills | | | | | |
| | | Problem-solving capacity | | | | | |
| | | Team/project leadership | | | | | |
| | | Developing, maintaining and using networks of collaborations | | | | | |
| | | Independent thinking/critical analysis/creativity | | | | | |
| | | Other skills, please specify: | | | | | |

| | | | |
|---|---|---|---|
| B.8 | *Going back in time, shouldn't you had the opportunity to join the activity on LHC, what would you have done as an alternative?*<br><br>*You can indicate more than one option* | ☐ I would have applied for a research period on another experiment at CERN<br>☐ I would have applied for a research period at another international research institute (other than CERN)<br>☐ I would have applied for a research period at another national research institute (other than CERN)<br>☐ I would have applied for a job in academia<br>☐ I would have applied for a job in the industry sector<br>☐ I would have applied for a job in the financial sector<br>☐ I would have applied for a job in the IT sector<br>Other, please specify: | |
| B.9 | *As of today, how many of the following have you authored/co-authored in relation to your research activity at LHC?*<br>*Please, indicate the total number for each of the following items* | Working papers/preprints: | |
| | | Articles in refereed journals: | |
| | | Papers for conference proceedings: | |
| | | Section/chapter in book: | |
| | | Patent: | |
| | | Software/application: | |
| | | Multimedia products: | |
| | | Other, please specify: | |
| B.10 | *Please indicate to which of the following experiments or research projects (other than the one indicated in question B.0), carried out at CERN, you have contributed for a period of at least one month.* | ACE<br>AEGIS<br>ALICE<br>ALPHA<br>AMS<br>ASACUSA<br>ATLAS<br>ATRAP<br>AWAKE<br>CAST<br>CLOUD<br>CMS<br>COMPASS<br>DIRAC | LHC machine<br>HiLumi LHC<br>LIU (Injector Upgrade)<br>ISOLDE<br>LHCb<br>LHCf<br>MOEDAL<br>NA61/SHINE<br>NA62<br>nTOF<br>OSQAR<br>TOTEM<br>UA9 |
| | | Please specify your activity: | |
| B.11 | *Please, indicate the total duration of your research period at CERN, including the activity indicated in question B.0 and others experiments/activities, if any.* | Number of months | |
| **YOUR CURRENT POSITION** | | | |
| | *What is your current position?* | I am still studying | **GO TO SECTION C** |
| | | I have completed my studies and I am currently unemployed | **GO TO SECTION C** |

|  |  | I am a post-doc researcher/I am working | GO TO SECTION D | | | |
|---|---|---|---|---|---|---|
|  | **PART C – YOUR EXPECTATIONS ABOUT YOUR PROFESSIONAL CAREER** | | | | | |
| C.1 | *Overall, to what extent do you expect that your experience at LHC will be relevant to your professional career? Please, provide an overall rate from 1 (= not relevant) to 5 (= very relevant).* | **1** | **2** | **3** | **4** | **5** |
| C.2 | *We would like to understand the impact that do you expect your experience at LHC will have on your salary. Please, answer to the following question by possibly thinking to somebody who has not been accepted at your experiment or other experiments at CERN.* | *To which extent do you expect that your future salary will be higher than that earned by somebody else?*<br><br>□ 0%<br>□ up to 5%<br>□ 5%, 10%<br>□ 11%, 20%<br>□ 21%, 30%<br>□ more than 30% | | | | |
| C.3 | *Please indicate your expectations about the <u>SECTOR</u> of your professional experience immediately after completing the studies.*<br>*In case you have completed your studies and you are currently unemployed, please indicate the SECTOR of your most desired professional experience*<br>*You can indicate more than one option* | □ Research (<u>at CERN</u>)<br>□ Research (<u>other than CERN</u>)<br>□ University<br>□ Other teaching<br>  Industry<br>  ICT sector (e.g. computing)<br>  Financial sector<br>  Public administration<br>□ Other, please specify | | | | |
| C.4 | *Please indicate the <u>POSITION</u> expected to be covered during your professional experience immediately after completing the studies.*<br>*In case you have completed your studies and you are currently unemployed, please indicate the <u>POSITION</u> of your most desired professional experience*<br><br>*You can indicate more than one option* | □ General Manager<br>□ Account Manager<br>□ Administrative/Office Manager<br>□ Finance manager<br>□ Software engineer<br>□ Mechanical engineers<br>□ Applications engineers<br>□ Engineering technician<br>□ Financial analyst<br>□ Data analyst<br>□ Physicist<br>□ Professor<br>  Associate Professor<br>  Lecturer/Assistant Professor<br>  Post doc<br>  Researcher<br>  Other, please specify | | | | |
| C.5 | **STARTING SALARY**<br><br>*What is the approximate <u>gross</u> annual salary do you expect to earn during your first professional experience (desired experience if you are currently unemployed)?* | □ less than 30,000 EUR<br>□ 30,000 – 40,000 EUR<br>□ 41,000 - 50,000 EUR<br>□ 51,000 - 60,000 EUR<br>□ 61,000 - 70,000 EUR<br>□ 71,000 - 80,000 EUR<br>□ 81,000 - 90,000 EUR<br>□ 91,000 - 100,000 EUR<br>□ 101,000 - 150,000 EUR<br>□ more than 150,000 EUR | | | | |

| | | |
|---|---|---|
| C.6 | **MID-CAREER SALARY**<br><br>*What is the approximate gross annual salary do you expect to earn in the mid of your career (desired experience if you are currently unemployed)?*<br><br>*Please, note that mid-career salary is the salary that you expect to earn after 10 years of working experience* | □ less than 30,000 EUR<br>□ 30,000 – 40,000 EUR<br>□ 41,000 - 50,000 EUR<br>□ 51,000 - 60,000 EUR<br>□ 61,000 - 70,000 EUR<br>□ 71,000 - 80,000 EUR<br>□ 81,000 - 90,000 EUR<br>□ 91,000 - 100,000 EUR<br>□ 101,000 - 150,000 EUR<br>□ more than 150,000 EUR |
| C.7 | **END-CAREER SALARY**<br><br>*What is the approximate gross annual salary that you expect to earn at the end of your professional career?* | □ less than 30,000 EUR<br>□ 30,000 – 40,000 EUR<br>□ 41,000 - 50,000 EUR<br>□ 51,000 - 60,000 EUR<br>□ 61,000 - 70,000 EUR<br>□ 71,000 - 80,000 EUR<br>□ 81,000 - 90,000 EUR<br>□ 91,000 - 100,000 EUR<br>□ 101,000 - 150,000 EUR<br>□ more than 150,000 EUR |

## PART D – YOUR PROFESSIONAL CAREER

**THE IMPACT OF YOUR EXPERIENCE AT LHC ON YOUR PROFESSIONAL CAREER**

| | | | | | | |
|---|---|---|---|---|---|---|
| D.1 | *Overall, to what extent your experience at LHC has been relevant to your professional career? Please, provide an overall rate from 1 (= not relevant) to 5 (= very relevant)* | 1 | 2 | 3 | 4 | 5 |

| | | | | | | | |
|---|---|---|---|---|---|---|---|
| D.2 | *Thinking to the possibility you had not joined the LHC, please indicate the extent to which you agree with the following statements about your PROFESSIONAL PATH.*<br>*Please, provide a rate from 1 (= strongly disagree) to 5 (= strongly agree) to each of the following items:* | *If I hadn't had the chance to join the LHC…* | 1 | 2 | 3 | 4 | 5 |
| | | *I would not have had difficulties to join an equivalent current job position* | | | | | |
| | | *I would have taken more time to find an equivalent current job position* | | | | | |
| | | *I would have been forced to look for job opportunities in other sectors outside the research sector* | | | | | |
| | | *Other, please specify:* | | | | | |
| D.3 | *Thinking to the possibility you had not joined the LHC, please indicate the extent to which you agree with the following statements about your SALARY.* | *If I hadn't had the chance to join the LHC …* | 1 | 2 | 3 | 4 | 5 |
| | | *I would have had a lower capacity to negotiate my salary and additional benefits with my employers* | | | | | |
| | | *I would have received the same salary* | | | | | |
| | | *I would have had few probabilities to increase my salary in the long-term* | | | | | |
| | | *Other, please specify:* | | | | | |

| | | | |
|---|---|---|---|
| D.4 | **We would like to understand the impact of your experience at LHC on your salary.** *Please, answer to the following question by possibly thinking to somebody who has not been accepted at your experiment or other experiments at CERN.* | *Thinking about that, to which extent is your CURRENT SALARY higher than that earned by somebody else.*<br><br>□ 0%<br>□ up to 5%<br>□ 5%, 10%<br>□ 11% , 20%<br>□ 21% , 30%<br>□ more than 30%<br><br>*Looking at your professional career in the long-term, to which extent, do you expect, your FUTURE SALARY will be higher than that earned by somebody else.*<br><br>□ 0%<br>□ up to 5%<br>□ 5%, 10%<br>□ 11% , 20%<br>□ 21% , 30%<br>□ more than 30% | |
| **FIRST CAREER MOVE** | | | |
| D.5 | *Please, indicate the year of start of your professional career* | List of Years | |
| D.6 | *Please, indicate the country of your first professional experience* | List of Countries | |
| D.7 | *With regard to your first professional experience, were you employed or associated at CERN?* | □ YES | **GO TO QUESTION D.7.1** |
| | | □ NO | **GO TO QUESTION D.8** |
| D.7.1 | *Please, specify your professional category at CERN* | Staff member<br>Apprentice<br>Fellow<br>User<br>Associate<br>Other, please specify<br><br>**GO TO QUESTION D.10** | |
| D.8 | *Please indicate the <u>SECTOR</u> of your first professional experience* | □ Research (<u>other than CERN</u>)<br>□ University<br>□ Other teaching<br>  Industry<br>  ICT sector (e.g. computing)<br>  Financial sector<br>  Public administration<br>□ Other, please specify | |

| | | |
|---|---|---|
| D.9 | *Please indicate the POSITION covered during your first professional experience* | □ General Manager<br>□ Account Manager<br>□ Administrative/Office Manager<br>□ Finance manager<br>□ Software engineer<br>□ Mechanical engineers<br>□ Applications engineers<br>□ Engineering technician<br>□ Financial analyst<br>□ Data analyst<br>□ Physicist<br>□ Professor<br>  Associate Professor<br>  Lecturer/Assistant Professor<br>  Post doc<br>  Researcher<br>  Other, please specify |
| D.10 | *What was the approximate gross annual salary earned during your first professional experience?* | □ less than 30,000 EUR<br>□ 30,000 – 40,000 EUR<br>□ 41,000 - 50,000 EUR<br>□ 51,000 - 60,000 EUR<br>□ 61,000 - 70,000 EUR<br>□ 71,000 - 80,000 EUR<br>□ 81,000 - 90,000 EUR<br>□ 91,000 - 100,000 EUR<br>□ *more than* 100,000 EUR |
| **YOUR CURRENT PROFESSIONAL EMPLOYMENT** | | |
| D.11 | *Is your current professional employment different from your first professional experience?* | □ YES **GO TO QUESTION D.12**<br>□ NO **GO TO QUESTION D.15** |
| D.12 | *Please indicate the SECTOR of your current experience* | □ Research (other than CERN)<br>□ University<br>□ Other teaching<br>  Industry<br>  ICT sector (e.g. computing)<br>  Financial sector<br>  Public administration<br>□ Other, please specify |
| D.13 | *Please indicate the POSITION covered during your current professional experience* | □ General Manager<br>□ Account Manager<br>□ Administrative/Office Manager<br>□ Finance manager<br>□ Software engineer<br>□ Mechanical engineers<br>□ Applications engineers<br>□ Engineering technician<br>□ Financial analyst<br>□ Data analyst<br>□ Physicist<br>□ Professor<br>  Associate Professor |
| | | Lecturer/Assistant Professor<br>Post doc<br>Researcher<br>Other, please specify |
| D.14 | *What is the approximate gross annual salary earned during your current professional experience?* | □ less than 30,000 EUR<br>□ 30,000 – 40,000 EUR<br>□ 41,000 - 50,000 EUR<br>□ 51,000 - 60,000 EUR<br>□ 61,000 - 70,000 EUR<br>□ 71,000 - 80,000 EUR<br>□ 81,000 - 90,000 EUR<br>□ 91,000 - 100,000 EUR<br>□ 101,000 - 150,000 EUR<br>□ more than 150,000 EUR |
| **YOUR EXPECTATIONS ABOUT THE FUTURE** | | |
| D.15 | *Do you expect to remain in the same SECTOR and to play the same POSITION in the future?* | □ YES  **GO TO QUESTION D.18**<br>□ NO   **GO TO QUESTION D.16** |
| D.16 | *Please indicate the expected SECTOR of your future career*<br>*You can indicate more than one option* | □ Research (at CERN)<br>□ Research (other than CERN)<br>□ University<br>□ Other teaching<br>  Industry<br>  ICT sector (e.g. computing)<br>  Financial sector<br>  Public administration<br>□ Other, please specify |
| D.17 | *Please indicate the expected POSITION of your future career*<br><br>*You can indicate more than one option* | □ General Manager<br>□ Account Manager<br>□ Administrative/Office Manager<br>□ Finance manager<br>□ Software engineer<br>□ Mechanical engineers<br>□ Applications engineers<br>□ Engineering technician<br>□ Financial analyst<br>□ Data analyst<br>□ Physicist<br>□ Professor<br>  Associate Professor<br>  Lecturer/Assistant Professor<br>  Post doc<br>  Researcher<br>  Other, please specify |
| D.18 | **MID-CAREER SALARY**<br><br>*What is the approximate gross annual salary do you expect to earn in the mid of your career?* | □ less than 30,000 EUR<br>□ 30,000 – 40,000 EUR<br>□ 41,000 - 50,000 EUR<br>□ 51,000 - 60,000 EUR<br>□ 61,000 - 70,000 EUR |

|  |  | *Please, note that mid-career salary is the salary that you expect to earn after 10 years of working experience* | □ 71,000 - 80,000 EUR<br>□ 81,000 - 90,000 EUR<br>□ 91,000 - 100,000 EUR<br>□ 101,000 - 150,000 EUR<br>□ more than 150,000 EUR |
|---|---|---|---|
| D.19 |  | **END-CAREER SALARY**<br><br>**What is the approximate <u>gross</u> annual salary that you expect to earn at the end of your professional career?** | □ less than 30,000 EUR<br>□ 30,000 – 40,000 EUR<br>□ 41,000 - 50,000 EUR<br>□ 51,000 - 60,000 EUR<br>□ 61,000 - 70,000 EUR<br>□ 71,000 - 80,000 EUR<br>□ 81,000 - 90,000 EUR<br>□ 91,000 - 100,000 EUR<br>□ 101,000 - 150,000 EUR<br>□ more than 150,000 EUR |